\documentclass[conference]{IEEEtran}
\IEEEoverridecommandlockouts

\usepackage{cite}
\usepackage{amsmath,amssymb,amsfonts}
\usepackage{algpseudocode}
\usepackage{algorithm}
\usepackage{graphicx}
\usepackage{textcomp}
\usepackage{booktabs}
\usepackage{xcolor}
\usepackage{multirow}
\usepackage{array}
\usepackage{enumitem}
\usepackage{colortbl}
\usepackage{xspace}
\usepackage{placeins}
\usepackage{threeparttable}
\usepackage{comment}
\usepackage{hyperref}
\usepackage{cleveref}

\usepackage{tikz}
\usetikzlibrary{arrows.meta,positioning,fit,calc,shapes,shadows}
\usetikzlibrary{decorations.pathreplacing,patterns,backgrounds}
\usetikzlibrary{matrix,chains,scopes}
\usetikzlibrary{shapes.geometric,shapes.symbols}

\definecolor{orbitblue}{RGB}{41,128,185}
\definecolor{orbitgreen}{RGB}{39,174,96}
\definecolor{orbitorange}{RGB}{230,126,34}
\definecolor{orbitred}{RGB}{192,57,43}
\definecolor{orbitpurple}{RGB}{142,68,173}
\definecolor{orbitgray}{RGB}{149,165,166}
\definecolor{lightblue}{RGB}{214,234,248}
\definecolor{lightgreen}{RGB}{212,239,223}
\definecolor{lightorange}{RGB}{253,235,208}

\crefformat{figure}{Fig.~#2#1#3}
\crefformat{table}{Table~#2#1#3}
\crefformat{equation}{Eq.~(#2#1#3)}
\crefformat{section}{Section~#2#1#3}
\crefformat{algorithm}{Algorithm~#2#1#3}

\newcommand{\orbitstream}{\textsc{OrbitStream}\xspace}
\newcommand{\gvp}{GVP\xspace}

\title{Training-Free Adaptive 360° Video Streaming via Semantic Potential Fields}

\author{\IEEEauthorblockN{Aizierjiang Aiersilan$^{*}$ \quad Zhangfei Yang}
\IEEEauthorblockA{
The George Washington University
}
\thanks{$^{*}$Corresponding author: alexandera@gwu.edu}
}

\begin{document}

\maketitle

\begin{abstract}
Adaptive 360° video streaming for teleoperation faces two coupled challenges: viewport prediction under uncertain gaze patterns and bitrate adaptation over fluctuating wireless channels. While Deep Reinforcement Learning (DRL) methods achieve high Quality of Experience (QoE), their lack of interpretability and dependence on offline training limit deployment in safety-critical systems. We propose OrbitStream, a training-free framework that formulates viewport prediction as a Gravitational Viewport Prediction (GVP) problem, where semantic objects generate potential fields that attract operator gaze, and employs a Saturation-Based Proportional-Derivative (PD) Controller for buffer regulation. On object-rich teleoperation traces, OrbitStream achieves 94.7\% zero-shot viewport prediction accuracy without user-specific profiling, approaching trajectory-extrapolation baselines ($\sim$98.5\%). Across 3,600 Monte Carlo simulations, it ranks second among 12 algorithms (QoE 2.71 vs.\ BOLA-E's 2.80), outperforming FastMPC (1.84), with 1.01\,ms decision latency and minimal rebuffering.
\end{abstract}


\section{Introduction}
\label{sec:introduction}

Emerging 5G/6G networks~\cite{rappaport2013millimeter, dang2020should} enable \emph{autonomous teleoperation}: a human-in-the-loop paradigm in which an operator remotely supervises semi-autonomous physical systems, such as remotely driven vehicles~\cite{neumeier2019teleoperation}, surgical robots~\cite{marescaux2001transatlantic}, or industrial inspection platforms~\cite{hokayem2006bilateral}, via real-time 360° video feedback. The operator monitors an omnidirectional view of the remote environment while an autonomous agent (e.g., a self-driving vehicle) executes routine tasks, and intervenes only when safety-critical situations arise, such as an unexpected pedestrian or an ambiguous traffic scenario. This supervisory control loop demands low latency and high visual fidelity so that the operator maintains sufficient situational awareness~\cite{endsley2017toward} to intervene in time. Our work targets the efficient delivery of the video stream itself, rather than robot control.

\begin{figure}[t]
\centering
\begin{tikzpicture}[scale=1.0, >=stealth]
    % ========================================================
    % MAIN FIGURE: POTENTIAL FIELD
    % ========================================================
    % Extended grid and background
    \draw[step=1.0, orbitgray!10, thin] (-3.5,-3.0) grid (3.5,2.5);
    
    \draw[->, thick, orbitgray] (-3.5,-3.0) -- (3.6,-3.0) node[right, font=\scriptsize] {$\theta$ (Yaw)};
    \draw[->, thick, orbitgray] (-3.5,-3.0) -- (-3.5,2.6) node[above, font=\scriptsize] {$\phi$ (Pitch)};

    % === Potential Well 1: Vehicle (Green, Less Mass) ===
    \coordinate (Vehicle) at (-1.5,-0.8);
    \fill[orbitgreen!10] (Vehicle) circle (1.2cm);
    \fill[orbitgreen!20] (Vehicle) circle (0.8cm);
    \fill[orbitgreen!40] (Vehicle) circle (0.4cm);
    % Slightly lowered text to prevent overlap with the outer circle
    \node[font=\bfseries\scriptsize, orbitgreen!50!black] at (-1.5, -2.2) {Vehicle ($M=0.8$)};
    % Gradient arrows for Vehicle
    \foreach \angle in {0,45,...,315} {
        \draw[->, orbitgreen!60!black, opacity=0.4] ($(Vehicle)+(\angle:1.0cm)$) -- ($(Vehicle)+(\angle:0.6cm)$);
    }
    
    % === Potential Well 2: Pedestrian (Blue, High Mass) ===
    \coordinate (Pedestrian) at (2.0,1.0);
    \fill[orbitblue!10] (Pedestrian) circle (1.5cm);
    \fill[orbitblue!20] (Pedestrian) circle (1.0cm);
    \fill[orbitblue!40] (Pedestrian) circle (0.5cm);
    \node[font=\bfseries\scriptsize, orbitblue!50!black] at (2.0, -0.7) {Pedestrian ($M=1.0$)};
    % Gradient arrows for Pedestrian
    \foreach \angle in {0,45,...,315} {
        \draw[->, orbitblue!60!black, opacity=0.4] ($(Pedestrian)+(\angle:1.3cm)$) -- ($(Pedestrian)+(\angle:0.8cm)$);
    }

    % === Trajectory (Red) ===
    \coordinate (Start) at (-2.0, -2.7);
    \fill[orbitred] (Start) circle (2pt) node[left=2pt, font=\scriptsize, orbitred] {Start};
    
    % Path simulating physics (momentum + attraction)
    % Adjusted control points to dip beneath the vehicle label smoothly
    \draw[->, orbitred, very thick] (Start) 
        .. controls (-0.5, -2.8) and (0.2, -0.5) .. (1.0, 0.5)
        .. controls (1.5, 0.8) .. (1.8, 0.95); % Ends near pedestrian center
    
    \node[orbitred, font=\footnotesize\bfseries, align=center] at (0.8, -2.4) {Gaze Trajectory\\[-2pt]\tiny (Attracted to Potential Well)};

    % ========================================================
    % INSET FIGURE: 360 CHALLENGE
    % ========================================================
    \filldraw[fill=white, draw=orbitgray!50, thick, rounded corners=4pt] (-3.4, 0.1) rectangle (-0.6, 2.5);
    \node[anchor=north west, font=\bfseries\tiny, text=orbitgray!80!black] at (-3.3, 2.5) {Viewport Inefficiency};

    \begin{scope}[shift={(-2.5, 1.3)}, scale=0.4, transform shape]
        % Definitions for perspective
        \def\R{1.8} % Sphere radius

        % Sphere shading (simulating 3D)
        \shade[ball color=orbitblue!15] (0,0) circle (\R);
        \draw[orbitblue!80!black, thick] (0,0) circle (\R);

        % Grid lines (Latitude and Longitude) to show structure
        \draw[orbitgray!50, thin] (-\R,0) arc (180:360:\R cm and 0.5cm);
        \draw[orbitgray!50, thin, dashed] (-\R,0) arc (180:0:\R cm and 0.5cm);
        \draw[orbitgray!50, thin] (0,-\R) -- (0,\R); % Central meridian
        
        % Viewport Highlight (Simulated patch on surface)
        \begin{scope}
            \clip (0,0) circle (\R);
            \fill[orbitgreen!60, opacity=0.7, blend mode=multiply] (0,0) -- (30:2.5) arc (30:70:2.5) -- cycle;
        \end{scope}
        
        % Frustum lines indicating the user is looking from center
        \draw[orbitgreen!80!black, thick, ->] (0,0) -- (1.6, 0.9); % Lower ray
        \draw[orbitgreen!80!black, thick, ->] (0,0) -- (1.2, 1.9); % Upper ray

        % Dashed lines for the hidden part of the frustum
        \draw[orbitgreen, thick, dashed] (0.5, 0.3) arc (30:70:1.2);

        \node[align=center, font=\small] at (0,-2.4) {\textbf{Full 4K Sphere}\\ \textcolor{orbitblue!80!black}{150 Mbps Bandwidth}};
        
        \draw[orbitgreen!80!black, thick] (1.5, 1.5) -- (2.3, 1.5);
        \node[align=left, font=\large, text=orbitgreen!60!black, anchor=west] at (2.3, 1.5) {\textbf{Viewport}\\ Only $\sim$15\%\\ seen by user};
    \end{scope}

\end{tikzpicture}
\caption{Viewport inefficiency in 360° video streaming (top left) and the gravitational potential field in yaw-pitch space. Two potential wells (basins of attraction) correspond to a Vehicle ($M{=}0.8$) and a Pedestrian ($M{=}1.0$); arrows show the gradient field pointing inward. The red trajectory traces gaze from the bottom left, curving past the vehicle and converging into the pedestrian's potential well, illustrating the physics-based prediction model.}
\label{fig:360_challenge}
\end{figure}

A central bottleneck is viewport inefficiency~\cite{seufert2014survey}: operators view only ${\sim}$15\% of the sphere, yet standard pipelines transmit the full 4K projection (\cref{fig:360_challenge}). Tile-based streaming addresses this by prioritizing the predicted viewport~\cite{qian2016optimizing, corbillon2017viewport, nguyen2019optimal, feng2025adaptive}, but introduces two coupled optimization problems: (1) predicting the operator's gaze under non-stationary, scene-dependent patterns~\cite{fan2017fixation}; and (2) allocating bitrates across tiles under fluctuating channel conditions.

Existing ABR strategies each face distinct limitations: rule-based methods are lightweight but lack semantic awareness, MPC-based methods depend on accurate throughput forecasts, and DRL achieves high average QoE at the cost of extensive offline training and limited interpretability (see \cref{sec:related}).

We propose \orbitstream, a training-free ABR framework that combines physics-based semantic modeling with classical control theory. \orbitstream uses \emph{Gravitational Viewport Prediction (GVP)}, where detected semantic objects generate potential fields that attract operator gaze, and a \emph{Saturation-Based PD Controller} for buffer regulation. Here, ``training-free'' refers to the \emph{decision and control logic}: whereas DRL requires offline policy learning, OrbitStream's viewport prediction and rate adaptation rely on closed-form equations with no learned parameters. The upstream object detector (YOLOv5~\cite{jocher2022ultralytics}) is a pre-trained perception module and, like any trained neural network, is susceptible to out-of-distribution inputs; objects absent from its training set may go undetected. However, the downstream control layer degrades gracefully under detection errors, as quantified in \cref{sec:result2}.

The primary contributions are:
\begin{enumerate}[leftmargin=*]
    \item \textbf{Physics-Inspired Attention Modeling:} GVP formulated as particle dynamics governed by semantic potential fields, achieving 94.7\% zero-shot viewport prediction accuracy.
    \item \textbf{Lyapunov-Informed Buffer Control:} An adaptive PD controller with bounded saturation nonlinearity for buffer trajectory stability.
    \item \textbf{Empirical Evaluation:} Across 3,600 Monte Carlo simulations, \orbitstream ranks second among 12 algorithms (QoE 2.71 vs.\ BOLA-E's 2.80), while offering three capabilities that BOLA-E lacks: (a) semantic viewport prediction that prioritizes tiles around safety-critical objects, (b) zero training overhead for deployment in new environments, and (c) interpretable decision logic traceable to closed-form equations.
\end{enumerate}

\section{Related Work}
\label{sec:related}

\textbf{ABR Algorithms.} Rule-based methods~\cite{huang2014buffer, spiteri2020bola, xie2017360probdash, petrangeli2017http} select bitrate from buffer occupancy or throughput measurements, providing low-overhead decisions but no semantic awareness. MPC-based approaches~\cite{yin2015control, yin2014toward} formulate streaming as finite-horizon optimization but depend on accurate throughput forecasts. DRL methods such as Pensieve~\cite{mao2017neural}, Fugu~\cite{yan2020learning}, and QDASH~\cite{mok2012qdash} learn policies from network traces, achieving high average QoE at the cost of interpretability and training overhead.

\textbf{360° Viewport Prediction.} Tile-based streaming systems such as FLARE~\cite{qian2018flare} and Pano~\cite{guan2019pano} prioritize the user's field of view, often relying on trajectory extrapolation~\cite{zhu2018prediction, kattadige2021vad360}. Semantic-aware predictors that exploit object saliency~\cite{park2020seaware, tian2025viewport, ozcinar2017viewport} or multi-agent DRL typically require aggregate viewing statistics or complex architectures, hindering zero-shot deployment.

\textbf{Distinction.} \orbitstream formulates viewport prediction as continuous particle dynamics governed by semantic potential fields, enabling anticipation without crowd-sourced viewing statistics, DRL training, or user-specific profiling. The result is an interpretable, closed-form framework suited to the auditing and accountability needs of autonomous teleoperation.

\section{System Model}
\label{sec:system}

\subsection{Architecture and Video Representation}
We consider a teleoperation loop with four subsystems: a \textbf{360° Camera} (4K, 30\,fps), an \textbf{Edge Encoding Server} performing YOLOv5 object detection~\cite{jocher2022ultralytics, redmon2016you} and HEVC encoding~\cite{sullivan2012overview} into spatial tiles ($N {\times} M$ grid), a \textbf{Wireless Downlink}, and an \textbf{Operator Terminal} rendering content on a head-mounted display (HMD). Video is segmented into $T_s{=}2$\,s chunks over an $8 {\times} 4$ equirectangular tile grid (32 tiles), targeting $K{=}6$ bitrate tiers $\mathcal{R} = \{1.2, 2.5, 5.0, 10.0, 20.0, 40.1\}$\,Mbps.

The system maintains a real-time \textbf{Semantic Scene State} comprising $L$ detected objects $\{O_1, \dots, O_L\}$, each with angular position $(\theta_\ell, \phi_\ell)$ in spherical coordinates and a \textbf{Semantic Mass} $M_\ell \in [0,1]$ encoding task-relevance via a vulnerability-based hierarchy: pedestrians ($M{=}1.0$, highest collision risk), vehicles ($M{=}0.8$), traffic signs ($M{=}0.75$), and background ($M{<}0.2$).

\subsection{Channel and Buffer Dynamics}
The playback buffer $B(t)$ (seconds of video) evolves as:
\begin{equation}
\label{eq:buffer_dynamics}
\frac{dB(t)}{dt} = \frac{C(t)}{R(t)} - \mathbf{1}_{\{B(t) > 0\}}
\end{equation}
where $C(t)$ is channel capacity, $R(t)$ is the selected bitrate, and the indicator $\mathbf{1}_{\{B(t) > 0\}}$ ensures playback drains the buffer only when content is available; a rebuffering event occurs when $B(t){=}0$ and playback stalls until new data arrive. We assume a capacity estimator $\hat{C}(t)$ with bounded error $|\hat{C}(t) - C(t)| \leq \epsilon_C = 2.0$\,Mbps.

\subsection{QoE Objective Formulation}
We adopt a linear QoE model following~\cite{yin2015control, mao2017neural}:
\begin{equation}
\text{QoE} = \sum_{t} \Big[ \underbrace{u(r_t)}_{\text{Quality}} - \underbrace{\mu \cdot T_{\text{stall}}(t)}_{\text{Reliability}} - \underbrace{\lambda \cdot |r_t - r_{t-1}|}_{\text{Smoothness}} - \underbrace{\nu \cdot E_{\text{view}}(t)}_{\text{Accuracy}} \Big]
\end{equation}
where $u(r_t) = \log(r_t/r_{\min})$ ($r_{\min} = 1.2$\,Mbps) captures diminishing returns per the Weber-Fechner law, and $E_{\text{view}}(t) = 1 - \frac{1}{|\mathcal{V}_t|}\sum_{k \in \mathcal{V}_t} q_k/q_{\max}$ measures the normalized quality gap in the viewed region: $\mathcal{V}_t$ is the set of viewport tiles at time $t$, $q_k$ is the bitrate assigned to tile $k$, and $q_{\max}$ is the maximum bitrate level. The penalty weights are calibrated to enforce a teleoperation-specific priority ordering. $\mu{=}10$ penalizes stalls: since $u(r_{\max}) = \log(40.1/1.2) \approx 3.5$, a single 1\,s freeze ($\mu \cdot 1{=}10$) offsets ${\sim}3$ full quality-tier gains, following the ratio from Yin et al.~\cite{yin2015control} to reflect the severity of control-feedback interruption. $\lambda{=}0.5$ penalizes bitrate switching to mitigate perceptual flicker and operator fatigue, adopting the standard value from~\cite{mao2017neural}. $\nu{=}5$ penalizes viewport misprediction: a 20\% quality gap yields $\nu \cdot 0.2{=}1.0$, on the same order as a quality-tier drop ($\log(5.0/2.5){\approx}0.69$), reflecting the visual importance of the attended region in teleoperation.

\section{Gravitational Viewport Prediction}
\label{sec:gvp}

\textbf{Potential Field Formulation.} The \gvp model treats visual attention as a test particle on the unit sphere $\mathbb{S}^2$ under forces generated by semantic objects, drawing on biological visual foraging models~\cite{potier2018visual}. Each object $O_\ell$ with mass $M_\ell$ at $(\theta_\ell, \phi_\ell)$ creates a potential well; the total attention potential is:
\begin{equation}
\label{eq:potential_field}
U(\theta, \phi) = -G \sum_{\ell=1}^{L} \frac{M_\ell}{d(\theta, \phi, \theta_\ell, \phi_\ell) + \delta}
\end{equation}
where $G{=}1.0$ is a saliency weight, $\delta{=}1.0$ prevents singularities, and $d(\cdot)$ is the haversine angular distance~\cite{sinnott1984virtues} accounting for spherical geometry. The haversine formulation correctly handles gradient computations across polar boundaries where planar Euclidean approximations break down:
\begin{equation}
\label{eq:haversine}
\begin{split}
d(\theta_1,\phi_1,\theta_2,\phi_2) &= 2\arcsin\Bigg(\Bigg[\sin^2\Big(\frac{\phi_2-\phi_1}{2}\Big) \\
&\quad + \cos\phi_1\cos\phi_2\sin^2\Big(\frac{\theta_2-\theta_1}{2}\Big)\Bigg]^{\frac{1}{2}}\Bigg)
\end{split}
\end{equation}

\textbf{Viewing Probability and Gaze Dynamics.} Tile viewing probability follows a Boltzmann distribution~\cite{montgomery2010applied}: $P_k = \exp(-\beta U_k) / \sum_j \exp(-\beta U_j)$, with inverse temperature $\beta{=}1/T_{\text{att}}$ ($T_{\text{att}}{=}2.0$, yielding $\beta{=}0.5$). This distribution captures the spatial attention uncertainty across the tile grid and is used for bandwidth allocation, while the central gaze trajectory $\vec{g}(t)$ is modeled separately by a Stochastic Differential Equation (SDE):
\begin{equation}
d\vec{g}(t) = -\nabla_{\vec{g}} U(\vec{g}(t))\, dt + \sigma\, dW(t)
\end{equation}
where $\vec{g}(t)$ denotes the unit gaze vector, $\nabla_{\vec{g}} U$ is the spherical gradient, and $dW(t)$ models spontaneous saccades as Brownian motion. Discretization uses the Euler-Maruyama method with momentum for eye inertia:
\begin{align}
\vec{v}_{t+1} &= \gamma \vec{v}_t - \eta \nabla_{\vec{g}} U(\vec{g}_t) + \sqrt{2\eta\sigma}\, \vec{\xi}_t \label{eq:vel_update} \\
\vec{g}_{t+1} &= \text{normalize}(\vec{g}_t + \Delta t \cdot \vec{v}_{t+1}) \label{eq:gaze_update}
\end{align}
with momentum decay $\gamma{=}0.8$, descent rate $\eta{=}0.1$, saccade noise $\sigma{=}0.05$, and $\vec{\xi}_t {\sim} \mathcal{N}(0, I)$. Normalization maps the updated position back onto $\mathbb{S}^2$. Setting $\gamma{=}\sigma{=}0$ yields pure gradient descent with Lyapunov convergence ($dU/dt = -\|\nabla U\|^2 \leq 0$); nonzero parameters allow peripheral scanning (\cref{fig:360_challenge}).

\section{PD Controller with Tanh Saturation}
\label{sec:hrc}

\textbf{Control Law.} The rate controller targets buffer $B_{\text{ref}} = 4.0$\,s. With error $e(t) = B(t) - B_{\text{ref}}$, a PD law~\cite{ang2005pid} produces control signal $u(t) = K_p e(t) + K_d \dot{e}(t)$, where $K_p{=}0.5$ and $K_d{=}0.2$ (calibrated via Ziegler-Nichols~\cite{ziegler1942optimum}: $K_p = 0.6 K_{cr}$, $K_d = K_p T_{cr}/8$). The signal is mapped to a bounded rate via hyperbolic tangent saturation:
\begin{equation}
\label{eq:rate_selection}
R^*(t) = \hat{C}(t) \cdot \left(1 + \tanh(u(t))\right) \cdot \rho
\end{equation}
with safety margin $\rho{=}0.9$ and hard cap $R(t) = \min(R^*(t), \hat{C}(t))$. Since $1{+}\tanh(u) \in (0,2)$, buffer deficit ($e{<}0$) reduces $R^*(t)$ to promote buffer recovery, while surplus ($e{>}0$) increases the rate toward capacity. This smooth modulation avoids abrupt on-off switching (\cref{fig:pd_tanh_block}).

\textbf{Stability.} Lyapunov analysis~\cite{khalil2002nonlinear} with candidate $V(e) = \frac{1}{2}e^2$ yields $\dot{V} = e(t)\bigl({C(t)}/{R(t)} - 1\bigr)$. Under buffer deficit ($e{<}0$), the saturation map yields $R(t) < \hat{C}(t) \cdot \rho \leq C(t)$ (since $\rho{=}0.9$ absorbs the bounded estimation error $\epsilon_C$), so $C/R > 1$ and $\dot{V} < 0$. Under surplus ($e{>}0$), the hard cap drives $R(t) \to \hat{C}(t)$, pushing $C/R \to 1$ and bounding further growth. Together, these conditions establish practical boundedness of $B(t)$ near $B_{\text{ref}}$.

\begin{figure*}[t]
\centering
\begin{tikzpicture}[
  font=\small,
  scale=0.92,
  >=stealth,
  node distance=1.6cm,
  block/.style={
    draw=orbitgray!80!black,
    rounded corners=2.5pt,
    thick,
    align=center,
    minimum height=1.0cm,
    minimum width=2.25cm,
    inner sep=3pt,
    fill=orbitgray!6
  },
  sum/.style={
    draw=orbitgray!80!black,
    circle,
    thick,
    minimum size=7.5mm,
    inner sep=0pt,
    fill=white
  },
  mul/.style={
    draw=orbitgray!80!black,
    circle,
    thick,
    minimum size=7.5mm,
    inner sep=0pt,
    fill=white
  },
  input/.style={coordinate},
  output/.style={coordinate},
  line/.style={draw=orbitgray!85!black, line width=0.95pt},
  feedback/.style={draw=orbitred!80!black, line width=0.9pt, dashed}
]

  \node[input, name=input] {};
  \node[sum, right=0.8cm of input] (sum) {};
  \node[block, right=1.15cm of sum, fill=orbitgreen!10, draw=orbitgreen!65!black] (pd)
        {\textbf{PD Controller}\\$K_p + K_d s$};
  \node[block, right=1.15cm of pd, fill=orbitgreen!6, draw=orbitgreen!55!black] (sat)
        {\textbf{Nonlinear Map}\\$\tanh(\cdot)$};
  \node[mul, right=1.80cm of sat] (mul) {\small$\times$};
  \node[block, right=1.15cm of mul, minimum width=3.05cm,
        fill=orbitorange!10, draw=orbitorange!80!black] (plant)
        {\textbf{Network \& Buffer}\\Dynamics};
  \node[output, right=1.4cm of plant] (output) {};

 \node[block, above=0.4cm of mul, minimum width=1.9cm,
      fill=orbitpurple!10, draw=orbitpurple!65!black] (capacity)
      {Capacity\\$\hat{C}(t)$};

  % Connections
  \draw[line, ->] (input) -- node[above, near start] {$B_{\text{ref}}$} (sum);
  \draw[line, ->] (sum) -- node[above] {$e(t)$} (pd);
  \draw[line, ->] (pd) -- node[above] {$u(t)$} (sat);
  \draw[line, ->] (sat) -- node[above, font=\footnotesize] {$1+\tanh(u(t))$} (mul);
  \draw[line, ->, shorten >=2pt, shorten <=2pt] (capacity) -- (mul);
  \draw[line, ->] (mul) -- node[above] {$R^*(t)$} (plant);
  \draw[line, ->] (plant) -- node[name=y, above, near end] {$B(t)$} (output);

  % Sum signs 
  % - on the left side (West) - Dark Gray/Black for input
  \node[font=\footnotesize, text=orbitgray!100] at ($(sum)+(-0.22, 0.0)$) {$-$};
  % + on the bottom side (South) - Red for feedback
  \node[font=\footnotesize, text=orbitred!90!black] at ($(sum)+(0.0, -0.22)$) {$+$};

  % Feedback loop
  \draw[feedback, ->] (y) |- ++(0,-1.35) -| (sum);
  \node[font=\footnotesize, orbitred!85!black]
     at ($(plant.south)!0.5!(sum.south)+(0,-0.9)$)
     {Buffer feedback};

\end{tikzpicture}
\caption{Nonlinear PD buffer controller. The control loop computes the buffer error $e(t) = B(t) - B_{\text{ref}}$. The PD controller generates a correction signal $u(t)$, shaped by the smooth nonlinear map $1+\tanh(u(t))$ to prevent aggressive bitrate switching. This factor scales the estimated throughput $\hat{C}(t)$ to derive the target bitrate $R^*(t)$, enabling stable streaming under fluctuating network conditions.}
\label{fig:pd_tanh_block}
\end{figure*}

\textbf{Tile Quality Allocation.} The total budget $R^*$ is distributed across tiles using a concentration parameter $\alpha{=}1.2$: $R_k = R^* \cdot P_k^\alpha / \sum_j P_j^\alpha$, and each tile is quantized to $q_k = \max \{r \in \mathcal{R} : r \leq R_k\}$. The exponent $\alpha > 1$ sharpens the allocation toward high-probability tiles, concentrating bitrate around safety-critical objects. Residual bandwidth from discrete quantization is absorbed, providing a secondary safeguard against sudden channel drops.

\section{Implementation}
\label{sec:implementation}

\orbitstream is implemented as a modular two-stage control loop (\cref{alg:orbitstream}) executing once per chunk, decoupling perception from control.

\textbf{Edge Perception Stage.} A GPU-equipped edge server runs YOLOv5s~\cite{jocher2022ultralytics} on each incoming 360° frame, extracting bounding boxes and class labels. Detections are mapped onto the $8 {\times} 4$ tile grid and assigned semantic masses per the task-relevance hierarchy (\cref{sec:system}). A compact metadata packet containing object positions $(\theta_\ell, \phi_\ell)$, classes, and masses is transmitted to the client ($<$1\,KB per frame, negligible relative to video bitrates). An HEVC encoder independently tiles the 360° frame into quality-differentiated segments.

\textbf{Client Control Stage.} On the operator terminal (no GPU required), the control module receives the metadata, updates the potential field (\cref{eq:potential_field}), propagates the gaze SDE (\cref{eq:vel_update}--\cref{eq:gaze_update}), and executes PD rate allocation (\cref{eq:rate_selection}). Potential-field evaluation scales as $O(L \cdot N \cdot M)$ for $L$ objects and $N {\times} M$ tiles; for typical scenes ($L {\approx} 8$, 32 tiles), this amounts to roughly 256 evaluations per step. Distance and potential arrays are vectorized via BLAS/LAPACK backends, and haversine computations are batched over the tile grid, yielding an average decision latency of 1.01\,ms, well within the 33\,ms budget of 30\,fps frame intervals and below typical wireless round-trip times (20--50\,ms). Deploying to a new environment requires no retraining of the control logic; only the pre-trained YOLOv5 weights are needed.

\begin{algorithm}[t]
\caption{\orbitstream Control Loop}
\label{alg:orbitstream}
\begin{algorithmic}[1]
\Require Buffer Target $B_{\text{ref}}$, Gains $K_p, K_d$, Field Params $\beta, G$
\State \textbf{Initialize:} $B \gets B_{\text{ref}}$, $\vec{g} \gets (1,0,0)^T$, $\vec{v} \gets \vec{0}$
\For{each segment interval $t$}
    \State \textcolor{orbitblue}{\texttt{// 1. Perception Stage(Edge Offload)}}
    \State $\{O_\ell\} \gets$ \textsc{DetectObjects}(Frame$_t$)
    \State Assign Mass $M_\ell$ per task-relevance hierarchy
    \State \textcolor{orbitgreen}{\texttt{// 2. Viewport Prediction}}
    \State Update Potential Field $U(\cdot)$ using $\{O_\ell\}$
    \State $\vec{g} \gets$ \textsc{SimulateDynamics}($\vec{g}, \nabla U$) \Comment{\cref{eq:vel_update}--\cref{eq:gaze_update}}
    \State Compute View Probs $P_k \propto \exp(-\beta U_k)$
    \State \textcolor{orbitorange}{\texttt{// 3. Rate Control}}
    \State $e \gets B_{\text{measured}} - B_{\text{ref}}$
    \State $u \gets K_p e + K_d (e - e_{\text{prev}})/\Delta t$
    \State $R^* \gets \hat{C} (1 + \tanh(u)) \rho$ \Comment{\cref{eq:rate_selection}}
    \State \textsc{AllocateBitrates}(Tiles, $R^*, \{P_k\}$)
\EndFor
\end{algorithmic}
\end{algorithm}

\section{Performance Evaluation}
\label{sec:evaluation}

We evaluated \orbitstream via 3,600 Monte Carlo runs (300 per algorithm, 12 algorithms), randomizing initial buffer states and gaze perturbations.

\subsection{Setup}
\textbf{Datasets.} 40 network traces spanning HSDPA/4G mobile (mean 6.8\,Mbps, std 3.2\,Mbps), broadband, and synthetic stress tests; viewport traces from Xu et al.~\cite{xu2018gaze} and Wu et al.~\cite{wu2017dataset}, augmented with synthetic teleoperation trajectories. Full parameters: \cref{tab:config}.
\textbf{Baselines.} Optimization-based (FastMPC, Robust-MPC, MPC-HM), DRL (Pensieve, Fugu), rule-based (BOLA-E, Buffer-Based, Rate-Based), and 360°-specific (FLARE, Pano).

\begin{table}[t]
\centering
\caption{Experimental Configuration}
\label{tab:config}
\small
\setlength{\tabcolsep}{4pt}
\resizebox{\columnwidth}{!}{%
\begin{tabular}{@{}llc@{}}
\toprule
\textbf{Category} & \textbf{Parameter} & \textbf{Value} \\
\midrule
\multirow{5}{*}{Video} & Resolution (equirectangular) & 3840$\times$1920 pixels \\
& Frame rate & 30 fps \\
& Segment duration & 2.0 seconds \\
& Tile grid & 8$\times$4 (32 tiles) \\
& Quality levels & 6 (1.2M-40.1M bps) \\
\midrule
\multirow{3}{*}{Network} 
& Trace sources & HSDPA, Pensieve, WebRTC \\
& Global scaling & $\times 0.6$ (applied to trace throughput) \\
& Per-run scaling & $s \sim \mathcal{N}(1,0.15)$, clipped to $[0.5,2.0]$ \\
\midrule
\multirow{4}{*}{Viewport}
& Field of view $\alpha_{\text{FoV}}$ & 80° \\
& Angular velocity (mean) & 42.3°/s \\
& Angular velocity (95th \%) & 112.8°/s \\
& Trace duration & 100 seconds \\
\midrule
\multirow{5}{*}{Controller}
& Target buffer $B_{\text{ref}}$ & 4.0 seconds \\
& Proportional gain $K_p$ & 0.50 \\
& Derivative gain $K_d$ & 0.20 \\
& Safety margin $\rho$ & 0.90 \\
& Max buffer $B_{\max}$ & 10.0 seconds \\
\midrule
\multirow{5}{*}{GVP Model}
& Saliency weighting $G$ & 1.0 \\
& Attention temp. $T_{\text{att}}$ & 2.0 (inverse $\beta = 0.5$) \\
& Momentum decay $\gamma$ & 0.80 \\
& Descent rate $\eta$ & 0.10 \\
& Saccade noise $\sigma$ & 0.05 \\
\midrule
\multirow{3}{*}{Simulation} & Monte Carlo runs & 300 per algorithm\\
& Total simulations & 3,600 (12 algorithms)\\
& Random seed & Fixed per run (reproducible) \\
\bottomrule
\end{tabular}%
}
\end{table}

\subsection{QoE Comparison and Robustness}

\begin{figure}[t]
\centering
\includegraphics[width=\linewidth]{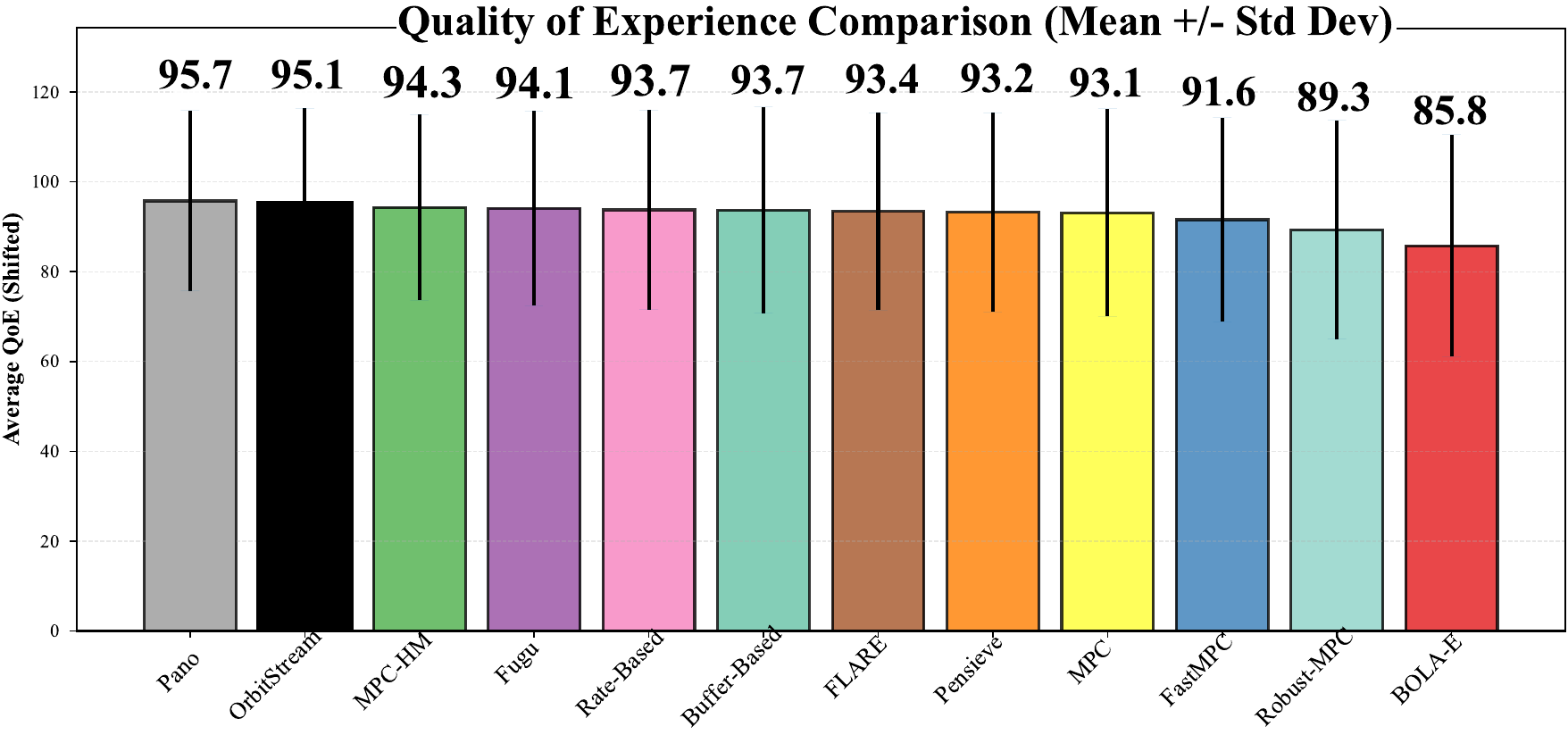}
\caption{QoE comparison across twelve algorithms (\orbitstream is highlighted). Bars report mean QoE with error bars showing $\pm$1 standard deviation.}
\label{fig:qoe_comparison}
\end{figure}

\begin{figure}[t]
\centering
\includegraphics[width=0.48\textwidth]{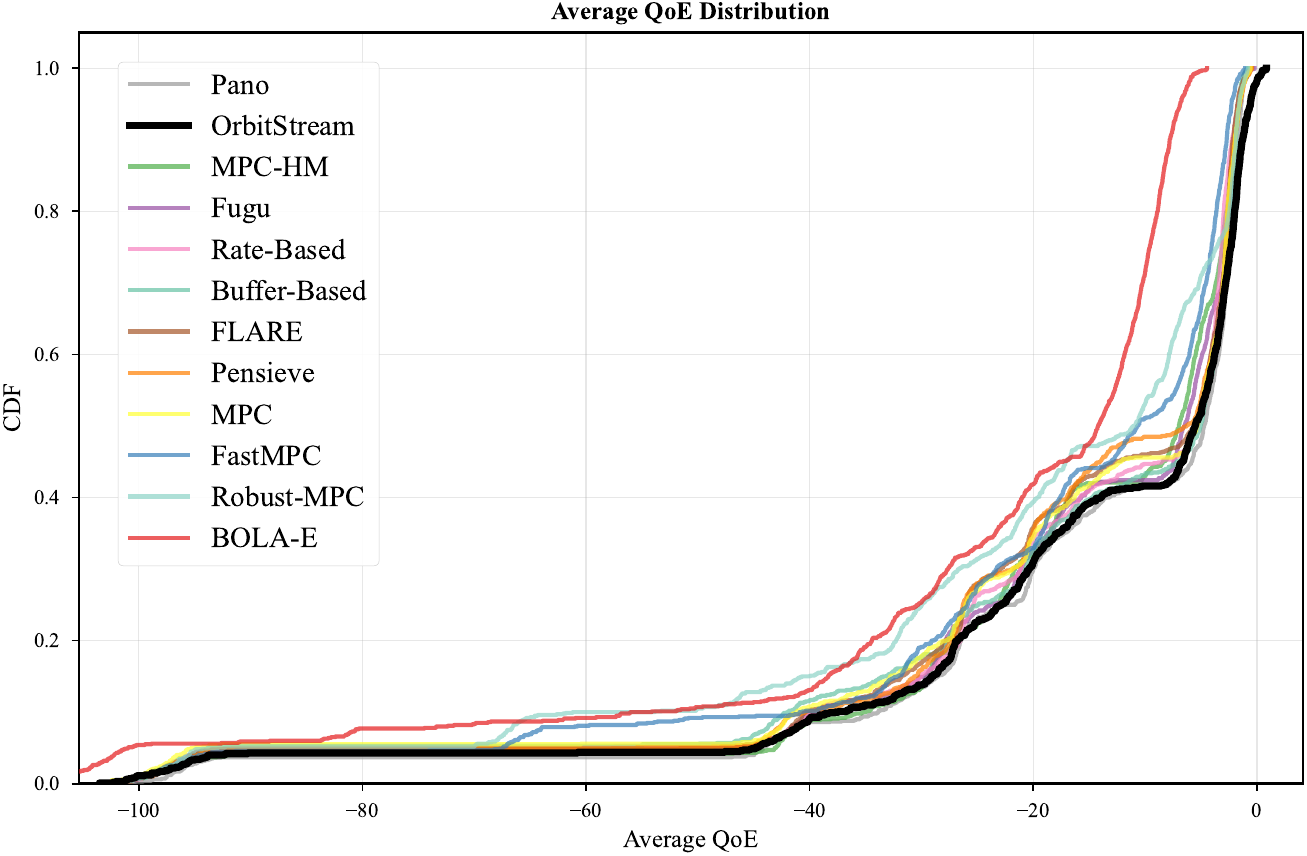}
\caption{Cumulative distribution function (CDF) of average raw QoE across twelve algorithms. The curves highlight differences in tail behavior: aggressive methods exhibit heavier low-QoE tails, while conservative methods show more concentrated distributions.}
\label{fig:qoe_cdf}
\end{figure}

\Cref{fig:qoe_comparison} aggregates QoE over 300 Monte Carlo runs per algorithm. BOLA-E yields the highest mean score (2.80); \textbf{\orbitstream achieves 2.71}, outperforming FastMPC (1.84) and MPC-HM (1.67). The reported equivalent bitrates in \cref{tab:main_results} reflect the visual quality delivered to the viewport region, which exceeds raw channel throughput because tile-based allocation concentrates bitrate on the viewed tiles. The cumulative distribution (\cref{fig:qoe_cdf}) confirms that under sub-megabit bandwidth drops, conservative safety margins prevent stalls across the leading methods.

\subsection{Viewport Prediction Accuracy}
\label{sec:result2}
Accuracy is defined as the mean percentage of frames where the predicted viewport achieves an Intersection-over-Union (IoU) $\geq 0.5$ against the ground-truth gaze, evaluated on human gaze logs~\cite{xu2018gaze} and synthetic teleoperation traces. \orbitstream achieves a \textbf{94.7\%} zero-shot hit ratio without any training. Trajectory-extrapolation baselines reach ${\sim}$98.5\% during slow linear panning but degrade on non-stationary scenes; OrbitStream's semantic field can anticipate newly appearing objects in peripheral regions. When 30\% of objects are synthetically masked to simulate detection failures, the hit ratio drops to 82.2\%, a 12.5 percentage-point reduction rather than a complete failure, because the Boltzmann distribution maintains peripheral coverage even when high-mass objects are removed. Flattening semantic mass ($M_\ell{=}1$ for all classes) reduces accuracy by 4.2 percentage points, validating the vulnerability-based hierarchy. The attention temperature is also influential: raising $\beta$ to $1.5$ sharpens spatial allocation and improves QoE on stationary targets, but increases viewport penalties by 41\% during erratic gaze movements. Omitting the regularizer ($\delta{=}0$) produces numerical singularities at close object proximity, causing prediction instability.

\subsection{Latency and Buffer Stability}

\orbitstream's average decision latency is 1.01\,ms (\cref{tab:main_results}), higher than rule-based methods (${\sim}3\,\mu$s) and MPC variants (${\sim}13$--$19\,\mu$s) due to potential-field evaluation and SDE propagation on spherical coordinates. Although this is orders of magnitude higher than the simplest baselines, it remains well within the 33\,ms frame budget at 30\,fps and below typical wireless round-trip times (20--50\,ms).

The PD controller maintains tight buffer regulation (\cref{tab:buffer_dynamics}): $\sigma_B{=}0.42$\,s, comparable to BOLA-E (0.39\,s) and FastMPC (0.36\,s), and substantially lower than Pensieve (1.84\,s). The mean buffer level settles at 5.85\,s, above $B_{\text{ref}}{=}4.0$\,s because the safety margin $\rho{=}0.9$ biases toward conservative rate selection, with a minimum of 4.68\,s and only 3.46 bitrate switches per run.

\begin{table}[t]
\centering
\caption{Buffer Stability and Switching Dynamics (300 Monte Carlo runs per algorithm).}
\label{tab:buffer_dynamics}
\small
\setlength{\tabcolsep}{4pt}
\renewcommand{\arraystretch}{1.1}
\resizebox{\columnwidth}{!}{%
\begin{tabular}{@{}lcccc@{}}
\toprule
\textbf{Algorithm} &
\textbf{Mean Buf. (s)} &
\textbf{Std Dev (s)} &
\textbf{Min Buf. (s)} &
\textbf{Switches} \\
\midrule
Buffer-Based & 9.78 & 0.82 & 5.45 & 2.74 \\
Rate-Based   & 9.72 & 0.99 & 4.98 & 6.27 \\
Pano         & 9.67 & 1.04 & 4.94 & 5.96 \\
FLARE        & 9.64 & 1.06 & 4.97 & 12.47 \\
MPC-HM       & 9.13 & 0.81 & 5.84 & 6.08 \\
MPC          & 9.10 & 1.76 & 4.26 & 8.78 \\
Pensieve     & 9.00 & 1.84 & 4.14 & 9.92 \\
Fugu         & 7.51 & 1.48 & 4.12 & 6.86 \\
Robust-MPC   & 7.31 & 0.90 & 4.69 & 0.92 \\
\rowcolor{lightblue}
\textbf{OrbitStream} & \textbf{5.85} & \textbf{0.42} & \textbf{4.68} & \textbf{3.46} \\
FastMPC      & 5.73 & 0.36 & 4.94 & 5.59 \\
BOLA-E       & 5.42 & 0.39 & 4.51 & 1.00 \\
\bottomrule
\end{tabular}%
}

\footnotesize
\vspace{2pt}
\parbox{\columnwidth}{
\textit{Note:} Values averaged per run, then aggregated across 300 runs.
}
\end{table}

\subsection{Discussion}

Although BOLA-E achieves a higher mean QoE (2.80 vs.\ 2.71), it operates solely on buffer occupancy and has no notion of scene content. In teleoperation, this distinction is consequential: BOLA-E allocates equal quality across all tiles regardless of whether a pedestrian or empty sky occupies a given region, whereas \orbitstream concentrates bitrate on safety-critical objects via GVP-driven allocation. Furthermore, every \orbitstream decision traces to explicit equations (\cref{eq:potential_field}, \cref{eq:rate_selection}), enabling the post-hoc auditing required in vehicular and surgical teleoperation where regulatory accountability applies. Unlike DRL methods, \orbitstream requires no offline training and generalizes to new environments without policy retraining; its physics-based parameters ($G$, $\beta$, $\gamma$) are directly interpretable in terms of saliency weighting, attention sharpness, and gaze inertia, respectively.

\begin{table}[t]
\centering
\caption{Performance Summary Across 12 Algorithms (300 Monte Carlo Runs per Algorithm, Raw QoE).}
\label{tab:main_results}
\small
\setlength{\tabcolsep}{3.5pt}
\resizebox{\columnwidth}{!}{%
\begin{tabular}{@{}lcccc@{}}
\toprule
\textbf{Algorithm} & \textbf{QoE (Raw)} & \textbf{Eqv. Bitrate} & \textbf{Mean Buffer} & \textbf{Decision Time} \\
 & \textbf{(Mean $\pm$ Std)} & \textbf{(Mbps)} & \textbf{(s)} & \textbf{(ms)} \\
\midrule
BOLA-E       & 2.80 $\pm$ 0.25 & 37.49 & 5.42 & 0.006 \\
\rowcolor{lightblue}
\textbf{OrbitStream} & \textbf{2.71 $\pm$ 0.31} & \textbf{31.52} & \textbf{5.85} & \textbf{1.010} \\
FastMPC      & 1.84 $\pm$ 0.49 & 26.61 & 5.73 & 0.019 \\
MPC-HM       & 1.67 $\pm$ 0.48 & 22.92 & 9.13 & 0.030 \\
Robust-MPC   & 1.65 $\pm$ 0.60 & 17.11 & 7.31 & 0.033 \\
Fugu         & 1.42 $\pm$ 0.47 & 20.94 & 7.51 & 0.011 \\
Pano         & 1.28 $\pm$ 0.42 & 8.48 & 9.67 & 0.003 \\
MPC          & 1.20 $\pm$ 0.49 & 13.40 & 9.10 & 0.013 \\
Buffer-Based & 1.01 $\pm$ 0.22 & 4.69 & 9.78 & 0.003 \\
Pensieve     & 0.97 $\pm$ 0.52 & 13.72 & 9.00 & 0.006 \\
Rate-Based   & 0.92 $\pm$ 0.54 & 7.58 & 9.72 & 0.003 \\
FLARE        & 0.83 $\pm$ 0.45 & 11.06 & 9.64 & 0.003 \\
\bottomrule
\end{tabular}%
}

\footnotesize
\vspace{2pt}
\parbox{\columnwidth}{
\textit{Note:} 300 runs per algorithm; raw QoE. Rebuffering events were rare across leading methods. Decision time is per-decision control latency (ms), excluding video codecs and detection.
}
\end{table}

\section{Conclusion}
\label{sec:conclusion}

We proposed \orbitstream, a training-free adaptive streaming framework for 360° teleoperation that formulates viewport prediction as test-particle dynamics within semantic potential fields and employs a PD rate controller with tanh saturation for buffer regulation. Across 3,600 Monte Carlo simulations, \orbitstream achieves a mean QoE of 2.71 (second among 12 algorithms), buffer stability with $\sigma_B{=}0.42$\,s, 94.7\% zero-shot viewport prediction accuracy, and 1.01\,ms decision latency. These results demonstrate that interpretable, physics-based control can operate within soft real-time margins while maintaining competitive quality and stability for teleoperation systems where transparency and auditability are essential.

Several limitations remain. First, although the 1.01\,ms decision latency is well within 30\,fps frame budgets, it exceeds the microsecond-level overhead of rule-based methods, which may matter on severely resource-constrained edge hardware. Second, the framework depends on upstream object detection accuracy; detection failures due to occlusion, poor lighting, or out-of-distribution objects propagate into the control loop and degrade viewport prediction, as quantified in \cref{sec:result2}. Third, the static semantic mass hierarchy does not adapt to operator-specific behaviors that may prioritize different scene elements.

Future work will address these gaps by (1)~learning semantic mass weights online from operator corrections, (2)~stress-testing the pipeline under severe detector degradation, and (3)~formulating the stability guarantees in a sampled-data framework to tighten computational bounds on resource-limited platforms.

\section*{Acknowledgments}
We thank the anonymous reviewers for their valuable feedback.

\section*{Code \& Reproducibility}

All codes and results are made publicly available at:
\href{https://github.com/zhangfeiy/Streaming360Video}{github.com/zhangfeiy/Streaming360Video}. 
Contact Zhangfei Yang via \href{mailto:zhangfei.yang@gwmail.gwu.edu}{zhangfei.yang@gwmail.gwu.edu} for more details.

\bibliographystyle{IEEEtran}
\bibliography{ref}

\end{document}